\documentclass[final,english]{bullsrsl}[2022/06/15]

\usepackage[latin1]{inputenc}
\usepackage[T1]{fontenc}

\usepackage{natbib} 
\usepackage{graphicx}

\begin{document}
\title{Onset and evolution of solar flares: 
Application of 2D and 3D models of magnetic reconnection
}

\author[affil={1}, corresponding]{Bhuwan}{Joshi}
\author[affil={2}]{Prabir K.}{Mitra}
\author[affil={2}]{Astrid M.}{Veronig}
\author[affil={1}]{R.} {Bhattacharyya}

\affiliation[1]{Udaipur Solar Observatory, Physics Research Laboratory, Udaipur 313001, India}
\affiliation[2]{University of Graz, Institute of Physics, Universitätsplatz 5, 8010 Graz, Austria}

\correspondance{bhuwan@prl.res.in}
\date{13th October 2020}
\maketitle



\begin{abstract}
The contemporary multi-wavelength observations have revealed various important features during solar flares which, on one hand, support the two-dimensional (2D) ``standard flare model'' while, on other hand, also urge for the exploration of three-dimensional (3D) magnetic field topologies involved in flares. Traditionally, the formation of parallel ribbons on both side of the polarity inversion line (PIL) and associated overlying loop arcades have been recognized as the most prominent features of eruptive flares which has formed the basis for the development of the standard model providing a 2D description of the flare-associated phenomena. The actual flare, however, occurs in a more complicated 3D magnetic structure. Thus, despite the general applicability, the standard model has limited or no scope in explaining some of the features which exclusively requires a 3D description. In this context, the observations of ``circular ribbon flares'' stand out where one of the ribbons presents an almost fully closed quasi-circular or quasi-ellipsoidal shape, evidencing the involvement of a typical fan-spine magnetic configuration. In this article, we discuss observational features vis-$\dot{a}$-vis theoretical understanding of solar flares in view of 2D and 3D models of magnetic reconnection. We highlight a few complex cases of circular ribbon flares exhibiting parallel ribbons, a coronal jet, and/or an erupting magnetic flux rope. Exploring various 3D topologies also enables us to probe similarities between the circumstances that govern the onset of jets, confined flares and CME-producing eruptive flares. 
\end{abstract}


\section{Introduction}
Solar flares are the most powerful explosions in our solar system, releasing energy up to 10$^{32}$-10$^{33}$ ergs in $\sim$10-1000~s. During a flare, energy that is stored in nonpotential magnetic fields in the solar corona is suddenly released through the process of magnetic reconnection.  Consequently, plasma in the corona and chromosphere gets heated to tens of million Kelvin. The heating results in striking enhancement of soft X-rays and longer-wavelength emissions \citep[for a comprehensive description of multi-wavelength flare observations, see reviews by][]{Fletcher2011,Benz2017}. At the primary energy release site, charged particles -- electrons, protons and heavier ions -- are accelerated to high energies and emit hard X-rays and $\gamma$-rays \citep[e.g.,][]{Ackermann2014,Karlicky2014}. Solar flares thus provide an unparalleled laboratory for investigating fundamental physical processes such as magnetic reconnection and the consequent particle acceleration and intense plasma heating. The extensive studies of solar flares have been possible due to the advent of various dedicated space-based solar missions in the last three decades: Yohkoh \citep{Ogawara1991}, SOHO \citep[Solar and Heliospheric Observatory;][]{Domingo1995}; TRACE \citep[Transition Region and Coronal Explorer;][]{Handy1999}; RHESSI \citep[Reuven Ramaty High Energy Solar Spectroscopic Imager;][]{Lin2002}; Hinode \citep{Kosugi2007}; SDO \citep[Solar Dynamics Observatory;][]{Pesnell2012}. 

Flares, mostly large ones, are frequently associated with the coronal mass ejections (CMEs). Together with CMEs, flares are known to drive hazardous space weather. Major space weather events can produce severe consequences for communication, power grids, aviation, spacecraft and other advanced technological systems -- making flare studies imperative with an ultimate goal toward space weather prediction.

Decades of solar flare studies have revealed that not all flares are associated with CMEs. In order to distinguish flares without accompanying CMEs from those that occur in association with CMEs, two distinct categories of flares -- ``confined'' and ``eruptive'' -- have been introduced. \cite{Yashiro2005} found that the probability of flares being associated with CMEs steeply increases with the X-ray flare magnitude. However, it should be noted that even large X-class flares can be of confined category. In this context, the flaring activity of NOAA active region (AR) 12192 is worth to mention. This AR produced a total of 6 X-class flares but none of them was associated with a CME \citep{SunX2015,Thalmann2015,Sarkar2018}. A common characteristic of many eruptive flares is prolonged soft X-ray emission which for large X-ray flares can extend upto several hours; such eruptive flares are known as ``long duration events (LDEs)''.

In this article, we summarize some important multi-wavelength aspects of solar flares. The typical observational characteristics of eruptive flares are discussed in Section~\ref{sec:standard-flare} where we also describe the standard flare model and its extended version. In Section~\ref{sec:3D-flare}, we discuss characteristics of solar flares in three-dimensional (3D) along with associated different magnetic topologies, particularly 3D null and quasi-separatrix layers (QSLs). 

\begin{figure}[t]
\centering
{\includegraphics[width=0.68\textwidth]{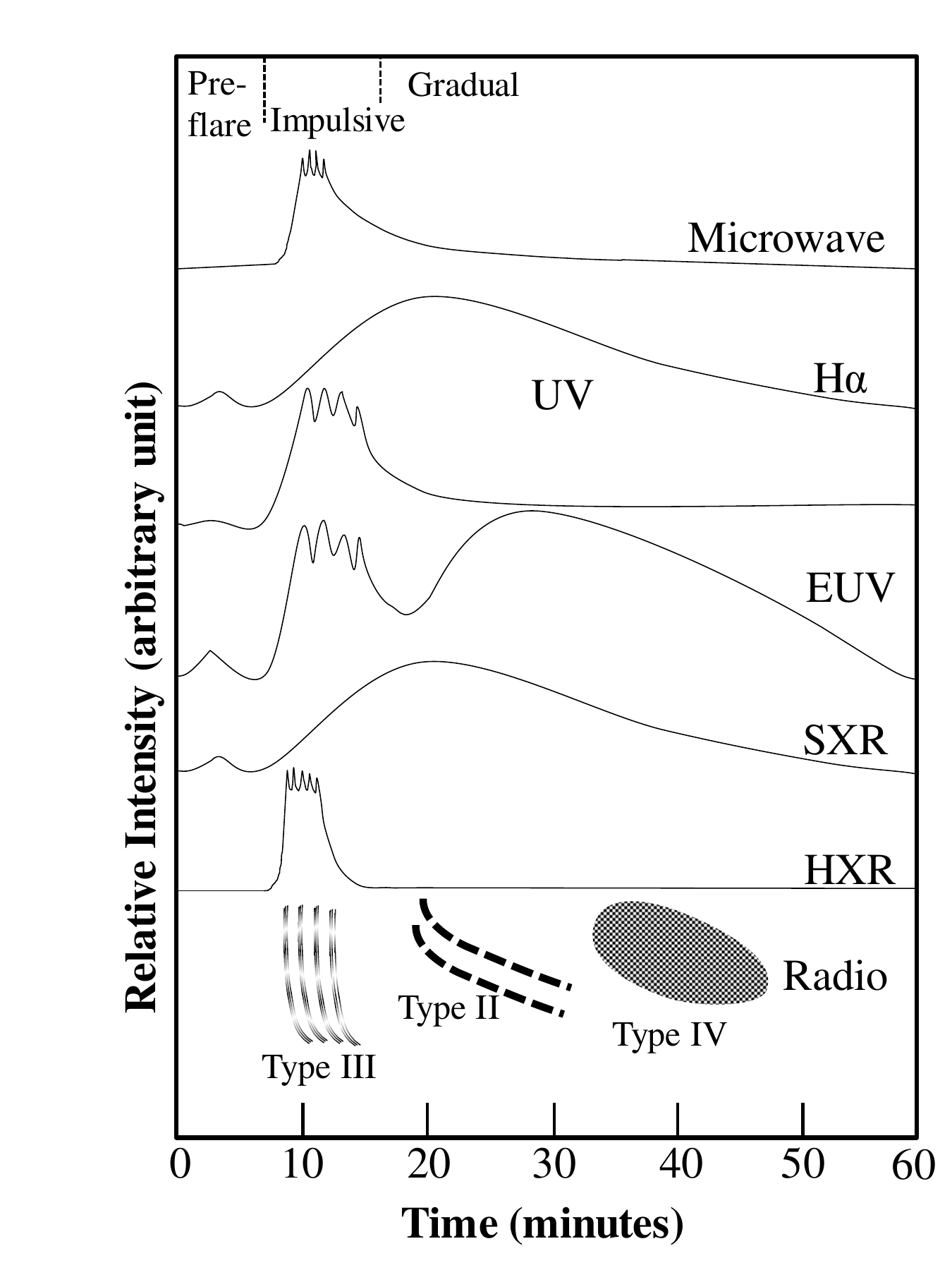}}
{\caption{Temporal evolution of a typical eruptive flare at different wavelengths.}\label{fig:temp_evo}}
\end{figure}

\section{``Standard'' flare: observations and models}
\label{sec:standard-flare}

\subsection{Multi-wavelength overview of two-ribbon flares}

During a flare, the released energy produces signatures over the entire electromagnetic spectrum, making it a truly multi-wavelength phenomenon. Flare radiations in different electromagnetic regimes are manifestations of various physical mechanisms taking place at different heights of the solar atmosphere, and therefore, they differ in time evolution. In Figure~\ref{fig:temp_evo}, we provide a schematic diagram to show the temporal evolution of a typical eruptive flare. Here we discuss observational signatures during the various phases of a solar flare.

\subsubsection{Preflare phase}
Prior to the explosive release of energy, many flares exhibit subtle yet noticeable changes in the neighbourhood of the main flare location, termed as the pre-flare phase. For clarity, we differentiate two types of phenomena during the pre-flare phase: {\it pre-flare activity} and {\it precursor emission}. The {\it pre-flare activity} is attributed to the very earliest stage which can also be temporarily disjoint from the onset of impulsive phase. These activities can include slow changes in the magnetic structures, viz. filament or hot channel activation, evolution of coronal loops, etc., but many times associated with only subtle enhancement in the emission \citep{Farnik1996,Farnik1998,Veronig2002,Harnandez2019,Mitra2020a}. On the other hand, the {\it precursor emission} is referred to the small-scale brightening readily observed in UV and SXR wavelengths, prior to the large-scale explosion during the impulsive phase \citep{Joshi2011,Joshi2016,Mitra2019}. The imaging observations further reveal that in many cases the distinct precursor events are not exactly co-spatial to the site in which bulk of flare radiation will subsequently originate but occur in the neighbourhood of it \citep{Farnik1996,Chifor2007}. We also note that some events exhibit metric type III radio bursts in the dynamic spectrum during the X-ray precursor emission which mark the earliest signatures of the magnetic reconnection \citep[e.g.,][]{Joshi2007,Mitra2019}.

\subsubsection {Impulsive phase}
\label{sec:impulsive}
\begin{figure}[t]
\centering
\includegraphics[width=\textwidth]{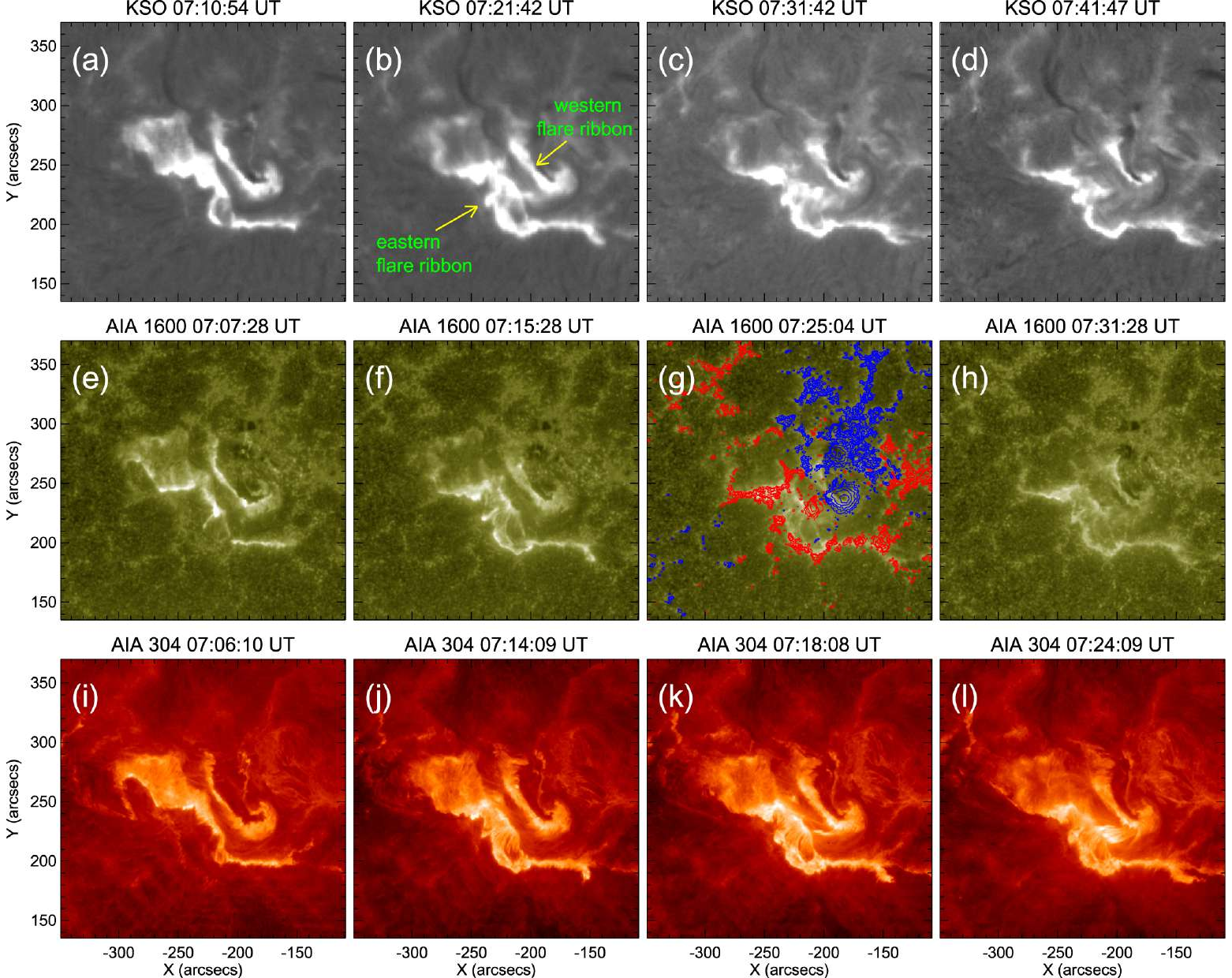}
\caption{Evolution of parallel ribbons during a typical eruptive flare (SOL2013-04-11T07:16, M6.5) shown in H$\alpha$ (top row), UV (middle row) and EUV (bottom row). The red and blue contours in panel (g) represent the distribution of positive and negative magnetic polarity, respectively. The H$\alpha$ observations shown here are from Kanzelh\"{o}he Observatory. The figure is adopted from \cite{Joshi2017a}.}
\label{fig:parallel_ribbon}
\end{figure}

\begin{figure}[t]
\centering
\includegraphics[width=\textwidth]{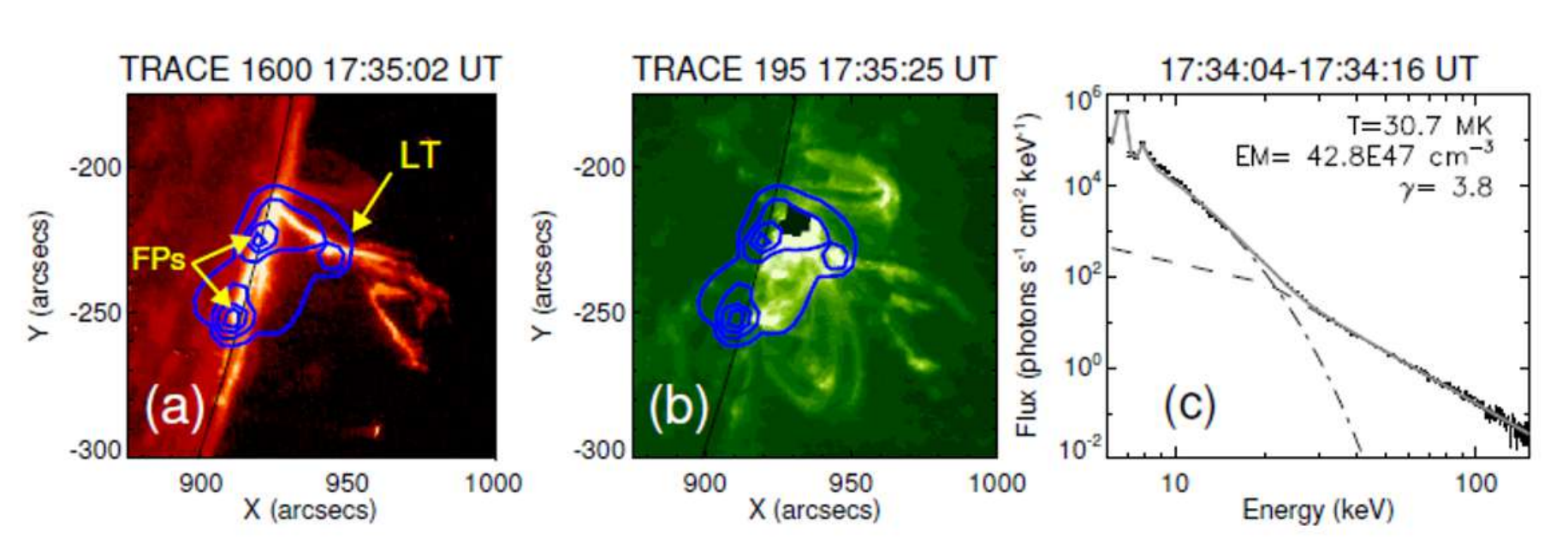}
\caption{Panels (a)--(b): Looptop (LT) and footpoints (FP) HXR sources (shown by blue contours) from a near-limb eruptive flare (SOL2004-08-17T17:40 (X1.8). Backgrounds in these images are UV and EUV images observed by TRACE. Panel (c): RHESSI X-ray spectroscopic analysis during the same event showing substantial heating of the solar atmosphere ($T$~$\sim$30~MK) as well as non-thermal emission with a hard power-law spectrum($\gamma < 4$). A detailed analysis of this event is presented in \cite{Cho2009,Joshi2013}.}
\label{fig:FT_LT}
\end{figure}

The bulk of the energy is released during the impulsive phase of the flare which lasts from tens of seconds to tens of minutes. The impulsive phase evidences the onset of large-scale coronal magnetic reconnection. This phase is characterised by emission in hard X-rays (HXR), non-thermal microwaves and in some cases also $\gamma$-rays and white-light continuum, showing evidence of strong acceleration of both electrons and ions. In metric radio wavelengths, the dynamic spectrum often exhibit type III radio busts, indicating the escape of relativistic elections along magnetic field lines opened to interplanetary space by coronal magnetic reconnection. The high energy non-thermal radiations during the impulsive phase are further supplemented by strong enhancement of emissions in chromospheric lines (e.g., H$\alpha$), ultraviolet (UV) and extreme ultraviolet (EUV). The impulsive phase radiations in H$\alpha$ and UV predominantly exhibit bright conjugate ``flare ribbons'' (also termed as ``parallel ribbons''; see Figure~\ref{fig:parallel_ribbon}) implying energy deposited to the chromospheric layers where the feet of the coronal loops that are subject to the flare are rooted at both sides of the magnetic polarity inversion line. In HXR observations, one or more compact sources are observed over each of the elongated ribbons called ``footpoints (FPs)'' sources (Figure~\ref{fig:FT_LT}). As the flare progresses, the two parallel flare ribbons (and HXR footpoints) separate from each other during the impulsive phase and later on toward the gradual phase \citep[e.g.,][]{Miklenic2007,Hinterreiter2018}. The HXR emission from the footpoints of flaring loops is traditionally viewed in terms of the thick-target bremsstrahlung process in which the X-ray production at the footpoints of the loop system takes place when high-energy electrons, accelerated in the coronal reconnection region, come along the guiding magnetic field lines and penetrate the denser transition region and chromospheric layers \citep{Brown1971}. An additional HXR source, called ``looptop (LP)'' source may appears right from the onset of the impulsive phase; which is easily distinguishable in limb flares (Figure~\ref{fig:FT_LT}). The HXR LT sources ascends to higher altitude and eventually disappears or fades away \citep[e.g.,][]{LinRP2003,Sui2004,Veronig2006,Joshi2007,Joshi2009,Joshi2016}. 

\subsubsection{Gradual phase} 

The gradual phase is most conspicuous in the SXR and EUV flare time profiles. During the impulsive phase, when the HXR and MW bursts dominate, the SXR and EUV emission gradually builds up in strength and peaks a few minutes after the impulsive emission. This gradual phase progresses with the formation of dense, bright loops (and arcade of loops in large flares), readily observed in the form of loop arcades in SXR and EUV images, indicating the presence of hot plasma ($\sim$10--20~MK) producing this emission. This hot plasma in the coronal loops comes from the chromosphere by a process termed as ``chromospheric evaporation'', introduced by \cite{Neupert1968}. In this process, the chromospheric plasma is first rapidly heated (at a rate faster than the radiative and conductive cooling rates) up to a temperature of $\sim$10$^{7}$ by the energy deposited by the energetic electrons which are accelerated at the magnetic reconnection site in the corona. The ensuing overpressure drives a mass flow upward along the overlying loops at a speed of a few hundred km~s$^{-1}$, which eventually fills the flaring loop with a hot plasma, giving rise to the gradual evolution of SXR emission \citep[e.g.,][]{Fisher1985}. This process is so striking that it results in a time delay between the peak in HXR (or microwave) and SXR time profiles of a flare, and the time derivative of the SXR emission exhibiting a similar profile like the HXR or microwave light curve \citep[e.g.,][]{Dennis1993,Veronig2005}. With the progression of coronal reconnection mainly in the impulsive phase (Section~\ref{sec:impulsive}), the new loops are formed at ever increasing altitudes while the older ones cool down. Thus, in the gradual phase of the flare, we find a well developed flare loop system which exhibits a gradient in temperature with outermost loops being the hottest ones. 

The radio dynamic spectra in meter wavelength often (but not always) shows type II and type IV radio bursts during the gradual phase \citep[see reviews by][]{Pick2008,Bastian1998}. Type II bursts, the consequence of plasma radiation associated with a magnetohydrodynamic shock propagating through the corona, are well correlated with eruptive flares \citep[e.g.,][]{Nelson1985,Cliver1999,Kumari2023}. The Type IV bursts are attributed to the emission from energetic electrons trapped either within an outward moving source (``magnetic cloud'' of a CME) or a stationary post-flare loop system \citep[e.g.,][]{Joshi2017a,Morosan2019,JoshiB2021}.

\subsection{Standard flare model}
\label{sec:standard-model}

The standard flare model successfully explains the origin of the most noticeable features of an eruptive flare -- flare loops and ribbons -- as a consequence of the large-scale magnetic reconnection in the corona. This model is essentially built on the collective works of \cite{Carmichael1964,Sturrock1966,Hirayama1974,Kopp1976} and, hence, is also called the CSHKP model. The standard flare model recognizes that the eruption of a filament or prominence distorts the overlying magnetic field configuration and stretches the field lines in such a way as that a vertical current sheet is produced beneath the erupting structure. Subsequently, magnetic field lines are subsequently dragged into this large-scale current sheet and reconnect to form the apparently growing flare loops and separating H$\alpha$ ribbons at their footpoints. Thus the key process is magnetic reconnection which releases sufficient magnetic energy on short time scales to account for the radiative and kinetic energies observed during an eruptive event. In this picture, the rise of the flare loop system (or the ascending HXR LT source observed in the impulsive phase) as well as the footpoint (or ribbon) separation reflect the upward motion of the magnetic reconnection site during  which field lines, rooted successively apart from the magnetic inversion line, reconnect. 

In Section \ref{sec:impulsive}, we have already introduced the FP and LT HXR sources in solar flares. In retrospect, let us briefly discuss how the X-ray observations of solar flares have immensely contributed in verifying the role of magnetic reconnection in solar flares which was proposed even before the two landmark X-ray observations of solar flares made by Japanese satellite Yohkoh: hot cusps and above-the-looptop HXR source. The observations from the Soft X-ray Telescope (SXT) onboard Yohkoh detected cusp-shaped structures above the hottest outer loops in eruptive LED flares \citep{Tsuneta1992}; the importance of this observations lies in the fact that the cusp resembles the general geometry of large-scale magnetic reconnection beneath the vertical current sheet and the retraction of newly reconnected filed lines toward the top of the already existing flare loops. Yohkoh made another landmark observations when its Hard X-ray Telescope (HXT) discovered an above-the-looptop source \citep{Masuda1994}. The observation of this new category of coronal source was a major breakthrough toward establishing the standard flare model as it provided the first observational evidences for a potential site of magnetic reconnection and electron acceleration in the corona. RHESSI observations further refined our knowledge about the X-ray emission from the coronal sources. Due to high sensitivity and broad energy coverage of RHESSI, the HXR emission from the looptop has now become a well known phenomenon and coronal HXR sources are detected in all phases of solar flares which are best seen and analysed in partially occulted flares \citep{Krucker2008,Effenberger2017}. However, we note that the above-the-looptop source is still a rarely observed feature \citep{Ishikawa2011}. 

Based on Yohkoh/SXT observations of blob-like hot plasma ejections, termed as ``plasmoid'' \citep{Shibata1998}, associated both with eruptive as well as confimed flares, \cite{Shibata1998a} introduced the {\it plasmoid-induced-reconnection model}, which is an extension of the CSHKP model. This model recognizes that the flare reconnection can be triggered by the eruption of a plasmoid (or a helical magnetic flux rope in 3D space). Thus, the eruption of a filament, plasmoid or magnetic flux rope produces alike consequences in stretching the overlying magnetic field lines which eventually triggers the magnetic reconnection.

\subsection{Failed eruptions}
Observationally, a particular variant of confined flares also involves so-called failed eruption of a prominence (or flux rope), where a flux rope (or prominence) is initially activated from the
source region and rises upward, but subsequently fails to escape from the overlying layers of the solar corona, and the material eventually falls back \citep[e.g.,][]{JiH2003,Gilbert2007,Kushwaha2015}. With an ever-increasing urge to understand the factors leading to CME
eruptions and develop methods to predict space weather, the observational and theoretical studies of failed eruptions have recently gained much attention and have become an important research topic in contemporary solar physics \citep[e.g.,][]{Sarkar2018,Amari2018, Mitra2022}. The physical scenario behind a failed or eruptive flares can be attributed to a competition between the ``hoop force" which assists the expansion of the magnetic flux rope and the tethering effect of the magnetic field lines overlying the rope \citep[e.g,][]{Torok2005,Amari2018}.

\section{Understanding the 3D nature of solar flares}
\label{sec:3D-flare}

Contemporary observations and numerical simulations have recently highlighted the three-dimensional (3D) nature of magnetic field lines and their reconnections pertaining to solar flares. With the standard flare model being essentially two-dimensional (2D), it is essential to focus on 3D magnetic topologies and their manifestations in solar flares -- or even more, coronal transients in general \citep[e.g., see][on the origin of extreme solar eruptive activity from NOAA 12673]{Joshi2022}. Observations that directly point toward the involvement of 3D magnetic field line distributions include shear in the flare loops \citep[e.g.,][]{Aulanier2012}; converging motions of flare ribbons and X-ray footpoint sources \citep[e.g.,][]{Joshi2009,Joshi2017a}; and J-shaped and circular ribbons \citep[e.g.,][]{Joshi2017a}, etc. In the context of 3D magnetic field topologies, {\it circular ribbon flares} have gathered much attention which we discuss in the next section.

\subsection{Circular ribbon flares, 3D magnetic nulls and quasi-separatrix layers}
\label{sec:CRF}
\begin{figure}
\centering
\includegraphics[width=0.8\textwidth]{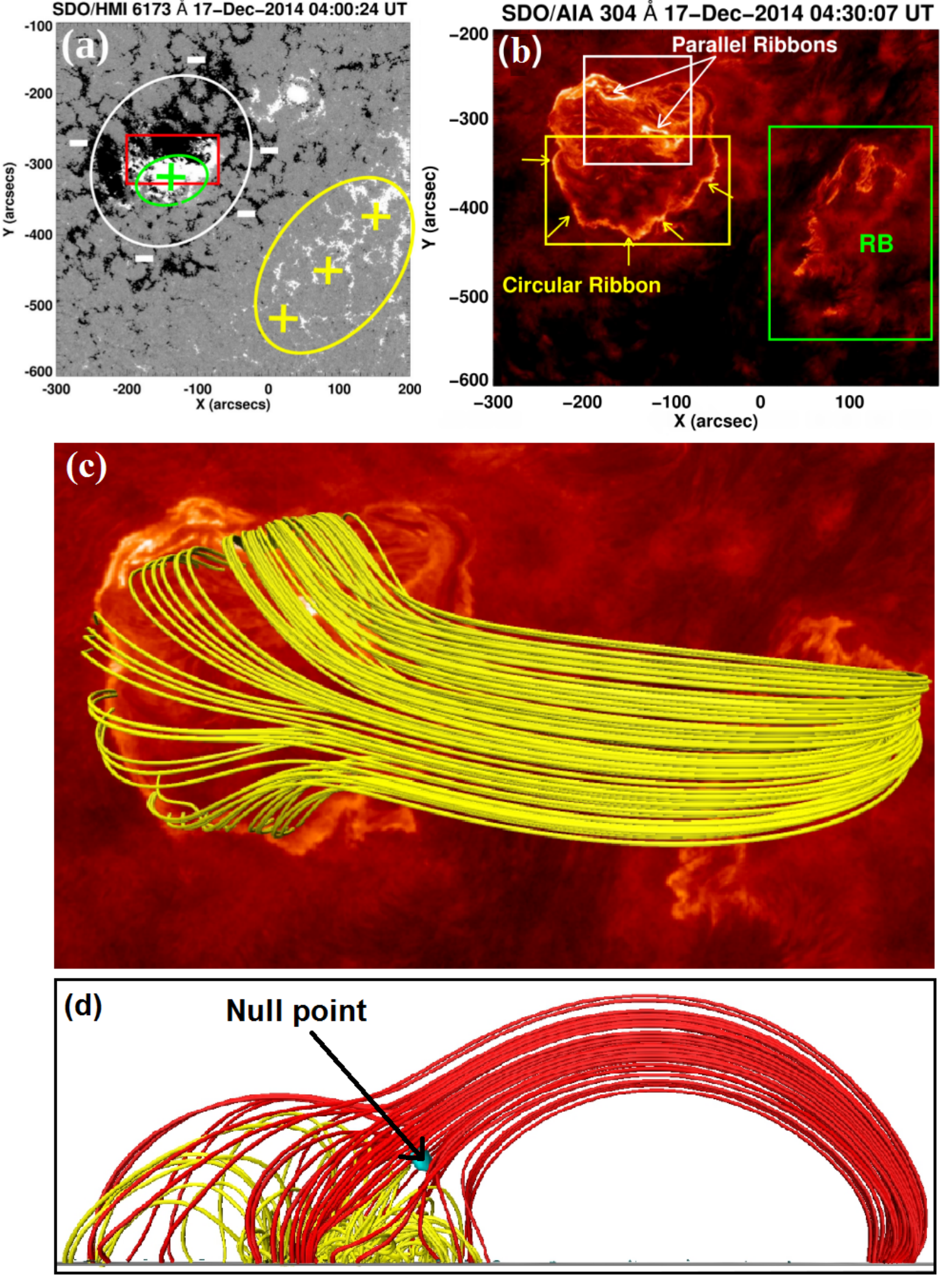}
\caption{Panel (a): an LOS HMI magnetogram of the active region NOAA 12242 showing typical magnetic configuration associated with circular ribbon flares. The central positive polarity and the surrounding negative polarity regions are represented by the green and white ellipse, respectively. Panel (b): AIA 304 \AA\ image showing the circular, parallel and remote brightenings from the same AR. The regions of circular, parallel and remote brightenings (RB) are enclosed by the yellow, white and green boxex, respectively. Panel (c): top view of the modeled fan-spine magnetic lines. Panel (d): side view of the modeled coronal magnetic configuration. The lines shown in panel (c) are shown in panel (d) by red color. The yellow lines in panel (d) represent another set of fan-spine lines. The location of the coronal null point between the two sets of fan-spine lines is represented by the blue contour. A detailed analysis of the magnetic topology of this active region in association with SOL2014-12-17T04:51 (M8.7) is presented in \cite{JoshiNC2021}.}
\label{fig:circular_ribbon}
\end{figure}

Recently a new subclass of flares have been recognized in which the ribbons display a quasi-circular or quasi-ellipsoidal shape \citep{Masson2009} instead of the traditional parallel ribbons on either side of the PIL during the impulsive phase. An example of such a circular ribbon flare (CRF) is shown in Figure~\ref{fig:circular_ribbon}. The CRFs can be
confined \citep[e.g.,][]{Hernandez-Perez2017,Devi2020,Cai2021}, as well as eruptive \citep[e.g.,][]{JoshiNC2015,JoshiNC2021}. Observations have also revealed jet and jets-like eruptions to be associated with CRFs \citep[e.g.,][]{Wang2012,Sahu2022,Nayak2019}. To further aid visualization of the relevant magnetic structures in 3D, we reproduce here the Figure~\ref{fig:circular_ribbon}(c)-(d) from \cite{JoshiNC2021} where the extrapolated magnetic field lines are overlaid on the photospheric magnetogram. 

3D magnetic reconnection can be sustained with various magnetic topologies. Here we focus on the topologies related to 3D magnetic nulls and quasi-separatrix layers because of their abundance in the solar corona \citep[see review by][]{Pontin2022}. 3D nulls are first and foremost points where magnetic field vanishes. To understand their magnetic topology, we generalize the notion of separatrix: the line in 2D which segregates sets of magnetic field lines having separate connectivities into three-dimensions. In 3D, the surfaces separating two  subvolumes of different field line connectivities are called separatrices. Notably, to maintain field line connectivities to be disjoint, the separatrices need to be also magnetic flux surfaces -- where magnetic field lines are tangential to the surface. If two such separatrices intersect, they intersect along a line throughout which the magnetic field has to be zero so as to make its magnitude unique. Conversely, the line joining two magnetic nulls is called a separator. The concept can be straightforwardly applied to magnetic structures in the solar corona. We often observe magnetic field structure in the photosphere where a parasitic polarity is emerged inside a larger opposite polarity region (Figure~\ref{fig:circular_ribbon}a). In such cases, the topological structure of a 3D null point defines a dome-like separatrix surfaces or separatrices, the fan, and two singular field lines, the spines, originating from the null point (Figure~\ref{fig:circular_ribbon}(c)-(d)). With the onset of reconnection at the null, magnetic field lines are transferred across the separatrices from one magnetic domain to another.

\cite{Priest1995} explored a way of generalizing the concept of separatrices of magnetic configuration without field-line linkage discontinuities. They put forward the idea that magnetic reconnection can also occur in 3D in the absence of null points at ``quasi-separatrix layers'' (QSLs), which are flat volumes where there is a rapid change in the field line linkage \citep[][]{Demoulin1997,Mandrini1997}. \cite{Titov2002} defined another characteristics function for QSLs: the Squashing degree $Q$. While the values of $Q$~$\gg$~2 correspond to QSLs, null-points are characterised by $Q$~$\rightarrow$~$\infty$.

\subsection{Various magnetic topologies and variants of circular ribbon flares (CRFs)}

\subsubsection{Fan--spine structure in CRFs and blowout coronal jets}
Comparison of the observed morphology of CRFs with corresponding magnetic field extrapolations reveal that the location of remote ribbons spatially matches with the remote footpoint of the outer spine field lines of the fan-spine topology while the circular ribbons trace the lower boundary of the filed lines making the fan-dome \citep[e.g.,][]{Devi2020, JoshiNC2021}. If the outer spine field lines are open outward (i.e., not terminating at a nearby region in the solar surface), the eruption along the outer spine may resemble a jet or jet-like activity \citep[e.g.,][]{Wang2012,Joshi2017b,Sahu2022}. Contextually, a data-based simulation by \cite{Nayak2019} has established magnetic reconnection at a 3D null having fan-spine structure is responsible for the generation of a coronal jet.

The above scenario further implies that the formation of remote ribbons is not guaranteed in all the cases and it esentially depends upon the connectivity of remote footpoints of spine field lines \citep[][]{Mitra2021,Mitra2023}. The study of \cite{Devi2020} shows a time delay of $\approx$2~min in the appearance of remote brightenings (which subsequently evolved into a remote ribbon) with respect to the appearance of the circular ribbon. The time delay in the appearance of the remote ribbon with respect to the circular ribbon has been noted in various events \citep{Wang2012,Li2018}. Based on the imaging analysis, \cite{Devi2020} supported the generally accepted idea that the most likely cause of the remote ribbon is the interaction of nonthermal particles, accelerated during the CRF and flowing along the overlying loops, with the dense chromospheric plasma at the distant footpoint of the overlying coronal loop system. 
The study of a confined CRF by \cite{Hernandez-Perez2017} points toward a different physical interpretation for the remote ribbons. Here the authors suggested that the remote brightenings are likely produced due to dissipation of kinetic energy of the plasma flows (i.e., heating due to compression); the flows are produced at the site of circular ribbon and are driven to the remote site along the overlying field lines. 

\begin{figure}[t]
\centering
\includegraphics[width=0.9\textwidth]{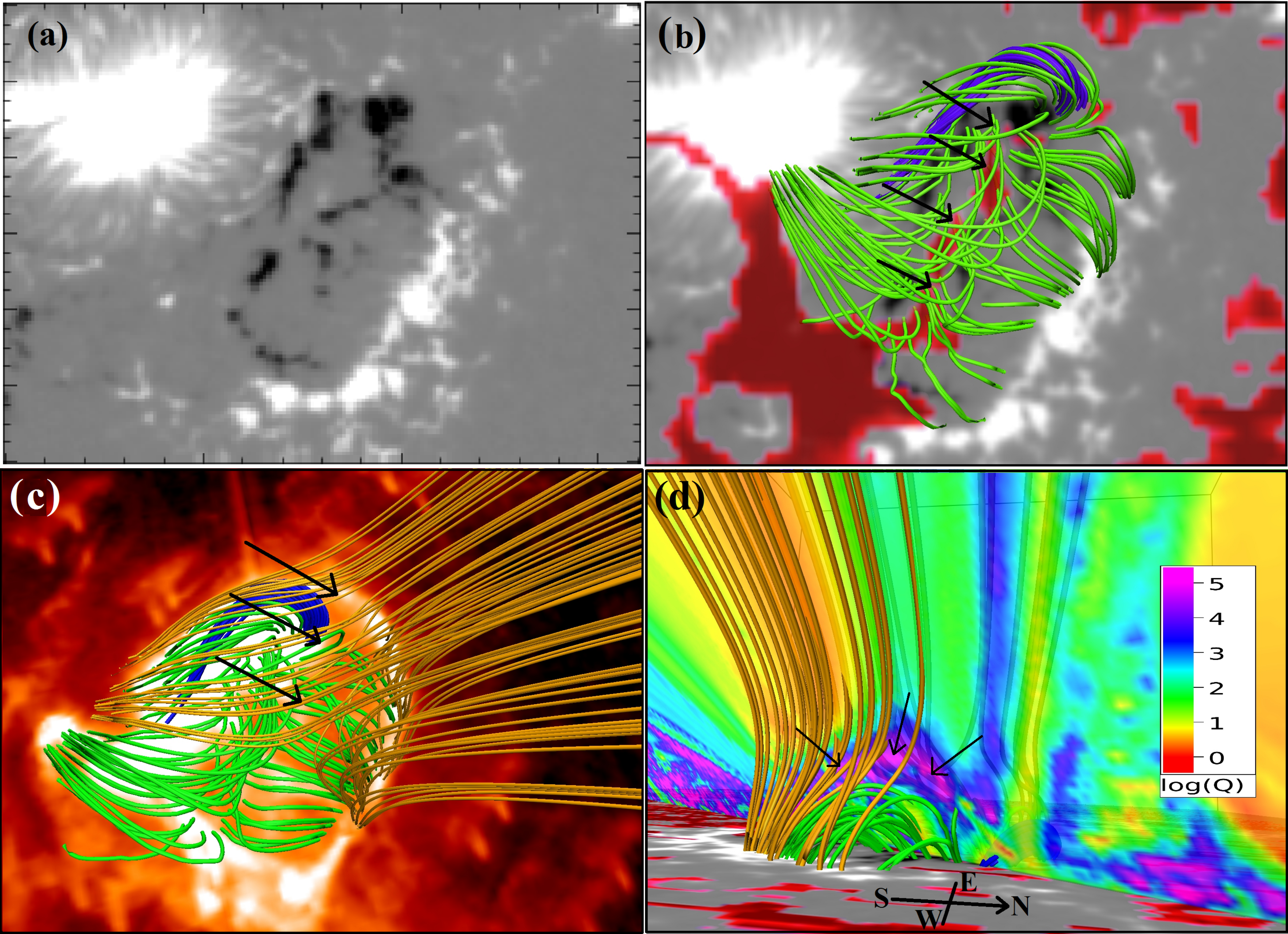}
\caption{Panel (a): LOS HMI magnetogram showing a magnetic atoll configuration in NOAA 11977 on 2014 February 11 reported by \cite{Mitra2021}. Panels (b)--(d): modeled coronal configuration associated with the atoll region. The blue, green and yellow lines represent a flux rope, the inner and outer fan-spine like configurations, respectively, The arrows in panels (b) and (c) demonstrate the laterally extended nature of the fan-spine-like filed lines in a configuration that lacked a coronal null point. Panel d: Distribution of the squashing factor $Q$ associated with the coronal magnetic configuration, where the HFT (indicated by the black arrows) can be identified. The red patches in panels (b) and (d) represent photospheric high-$Q$ regions (log($Q$)~>~2).}
\label{fig:HFT}
\end{figure}

\subsubsection{Parallel ribbon within large-scale circular ribbons}
Complex cases of the formation of classical parallel flare ribbons inside the periphery of large-scale circular ribbons have been reported (e.g. Figure~\ref{fig:circular_ribbon}(b)) in several studies \citep{JoshiNC2015, Devi2020, JoshiNC2021}. The first observation of this kind was reported by \cite{JoshiNC2015} where the authors found the existence of a large-scale fan-spine type magnetic configuration with a sigmoid lying under a section of the fan dome. Based on their detailed multi-wavelength analysis and coronal magnetic field modeling, a two step mechanism was proposed: the reconnection occurring in the wake of the erupting sigmoid produces the parallel flare ribbons on both the sides of the circular polarity inversion line; afterwards, the null-type reconnection higher in the corona, possibly triggered by the erupting sigmoid, leads to the formation of a large quasi-circular ribbon.

\subsubsection{Homologous quasi-CRFs involving a hyperbolic flux tube} 
\cite{Mitra2021} reported a unique case of four homologous quasi-circular ribbon flares that were triggered by the erupting filaments in AR NOAA 11977 during an interval of $\sim$11 h. The flaring region was associated with a a unique fan-spine-like configuration that developed over a complex photospheric configuration where dispersed negative polarity regions were surrounded by positive polarity regions (Figure~\ref{fig:HFT}). This unique photospheric configuration resembles the geological `atoll' shape. Computation of the degree of squashing factor in the region, i.e. $Q$-map, clearly revealed an elongated region of high $Q$-values between the inner and outer spine-like field lines, implying the presence of an hyperbolic flux tube (HFT). During the interval of the four flares, the authors observed a continuous decay and cancellation of negative polarity flux within the atoll region. Accordingly, the apparent length of the HFT gradually reduced to a null-point-like configuration before the fourth flare.

\begin{figure}[t]
\centering
\includegraphics[width=0.9\textwidth]{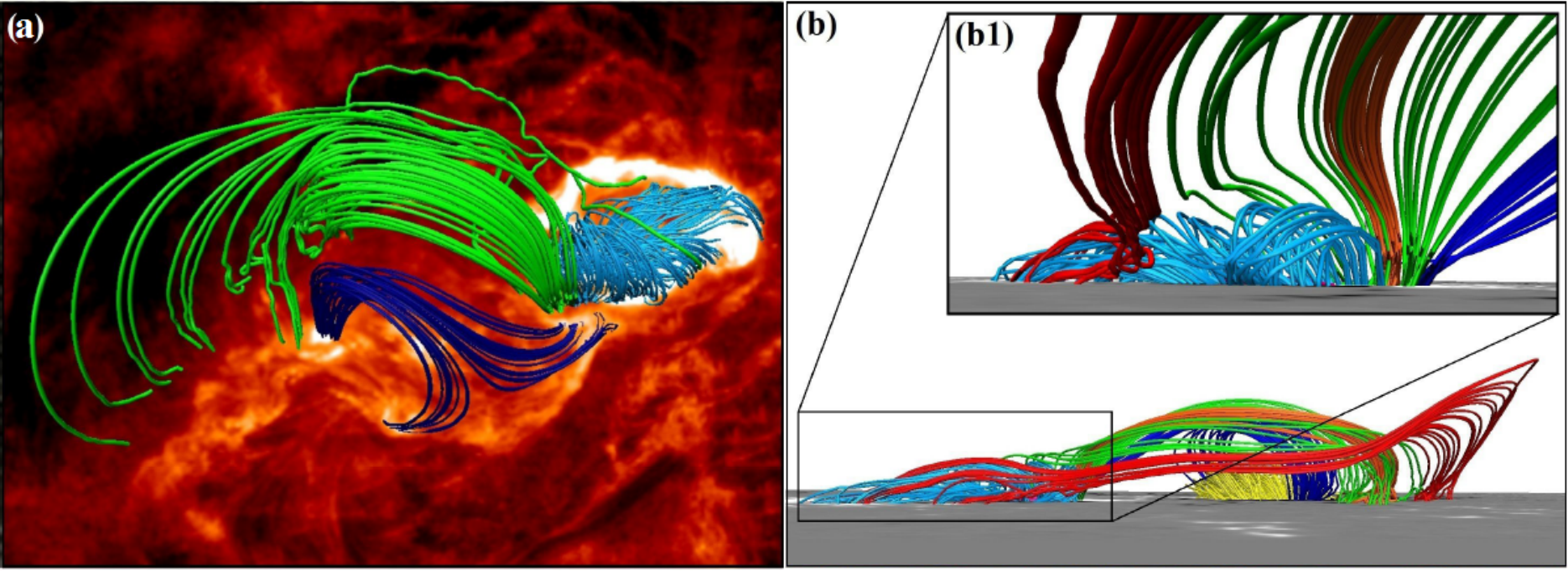}
\caption{A circular ribbon flare (SOL2015-03-12T21:51, M2.7) with incomplete fan-spine configuration reported by \cite{Mitra2023}. The incompleteness of the fan-spine configuration is shown from different angles in panels (b) and (b1) for better visualisation.}
\label{fig:incomplete_FS}
\end{figure}

\subsubsection{CRFs triggered from an incomplete fan-spine configuration}
\cite{Mitra2023} present observations of a complex circular ribbon class
flare which occurred in an`incomplete fan-spine-like' configuration that did not
manifest a coronal null point. The photospheric configuration associated with the flare differs from classical anemone-type active regions, as the central negative-polarity region is surrounded by positive-polarity regions on only three sides, that is, it is not completely surrounded (Figure~\ref{fig:incomplete_FS}). Using RHESSI X-ray observations the authors identified clear X-ray sources (up to 25 keV) from the footpoints of the QSL structure, which suggests that slipping reconnection can also lead to discernible signatures of particle acceleration. 

\section{Conclusion}
In this article, we have reviewed various observational aspects of solar flares. The standard 2D flare model broadly covers morphology and kinematics of an eruptive flare but it has limitations in explaining many features observed during the pre-flare phase and main flare as those features predominantly involve 3D magnetic structure. In the context of 3D magnetic configurations, the observations of circular ribbon flares are particularly striking which we have discussed with various complex examples. Nevertheless, understanding configurations and reconnections in 3D is challenging and leaves plenty of room for discussions and investigations. Future works pertaining to the study of various magnetic field topologies in the solar active regions through observations and simulations will provide us a deeper understanding of magnetic reconnection in 3D.

\begin{furtherinformation}

\begin{orcids}
\orcid{0000-0001-5042-2170}{Bhuwan} {Joshi}
\orcid{0000-0002-0341-7886}{Prabit K.}{Mitra}
\orcid{0000-0001-5042-2170}{Astrid M.}{Veronig}
\orcid{0000-0003-4522-5070}{R.}{Bhattacharyya}

\end{orcids}

\begin{authorcontributions}
BJ wrote the first draft and lead the discussions with the co-authors. PKM prepared figures. All the authors have read the manuscript and enriched it with their comments and suggestions.
\end{authorcontributions}

\begin{conflictsofinterest}
The authors declare no conflict of interest.
\end{conflictsofinterest}

\end{furtherinformation}

\bibliographystyle{bullsrsl-en}

\bibliography{extra}

\end{document}